# Quantum-Enhanced Polarimetric Imaging

Meng-Yu Xie,[1, 2, †] Su-Jian Niu,[1, 2, 3, †] Zhao-Qi-Zhi Han,[1, 2] Yin-Hai Li,[1, 2] Ren-Hui Chen,[1, 2] Xiao-Hua Wang,[1, 2] Ming-Yuan Gao,[1, 2] Li Chen,[1, 2] Yue-Wei Song,[1, 2] Zhi-Yuan Zhou,[1, 2, *] and Bao-Sen Shi[1, 2, *]

1 Key Laboratory of Quantum Information, University of Science and Technology of China, Hefei, Anhui 230026, China

2 Synergetic Innovation Center of Quantum Information & Quantum Physics, University of Science and Technology of China, Hefei, Anhui 230026, China

3 Xinjiang Key Laboratory for Luminescence Minerals and Optical Functional Materials, School of Physics and Electronic Engineering, Xinjiang Normal University, Urumqi, Xinjiang 830054, China

† These two authors contribute equally to this article

* Corresponding authors: drshi@ustc.edu.cn; zyzhouphy@ustc.edu.cn

**Abstract:** Polarimetric imaging, a technique that captures the invisible polarization-related properties of given materials, has broad applications from fundamental physics to advanced fields such as target recognition, stress detection, biomedical diagnosis and remote sensing. The introduction of quantum sources into classical imaging systems has demonstrated distinct advantages, yet few studies have explored their combination with polarimetric imaging. In this study, we present a quantum polarimetric imaging system that integrates polarization-entangled photon pairs into a polarizer-sample-compensator-analyzer (PSRA)-type polarimeter. Our system visualizes the birefringence properties of a periodical-distributed anisotropic material under decreasing illumination levels and diverse disturbing light sources. Compared to the classical system, the quantum approach reveals the superior sensitivity and robustness in low-light conditions, particularly useful in biomedical studies where the low illumination and non-destructive detection are urgently needed. The study also highlights the nonlocality of entangled photons in birefringence measurement, indicating the potential of quantum polarimetric system in the remote sensing domain.

**Teaser:**
A two-dimensional quantum polarimetric imaging system that can accurately and remotely image the invisible birefringence distribution at extremely low light intensities.

## Introduction

Polarization, one of the fundamental properties of light, exhibits various physical features of the tested specimen that are invisible to the human eye. The polarimetric imaging technique is used to visualize these polarization-related properties of the imaged objects or media. It can be used to enhance the contrast between the background and the target, which is difficult to distinguish in conventional imaging (1). It also provides a way to measure invisible parameters undetectable by conventional imaging, such as optical constants (2), chiral properties (3), and stress/strain distributions (4, 5). Nowadays, polarimetric imaging has been widely used for a variety of practical scenarios, such as 3D shape reconstruction (6, 7), solar and atmospheric phenomena (8), terrestrial vegetation and ocean detection (9, 10), biomedical diagnosis (11, 12), navigation (13) and so on.

The introduction of a quantum light source into conventional imaging opens up a new way to enhance the measurement accuracy especially in low illumination fields (14, 15). The ability to break the standard quantum limit has been substantiated both theoretically and experimentally, demonstrating great promise for the field of metrology (16, 17). Furthermore, the nonlocal properties of the entangled photon source facilitate the separation of the control and measurement processes (18), a feature of great value in the field of remote sensing applications. So far, the mainstream of the quantum imaging work focuses on accurately measuring the external morphology of the tested sample (19). The "invisible" polarization-correlated quantum imaging has only a limited amount of research, and most of them are single-spot detection of the homogeneous materials (20-22). The works combining the polarimetric imaging and quantum imaging are extremely rare (23), and the corresponding researches are still remained to be explored.

In this work, we introduce polarization entangled photon pairs into a polarizer-sample-compensator-analyzer (PSRA)-type polarimeter to form a two-dimensional quantum polarimetric imaging system. An anisotropic material with periodically distributed birefringence properties is selected as the sample to be imaged under diminishing illumination intensity and different sources of disturbance light. By comparing the classical and quantum polarimeters under the same conditions, we use the structural similarity index (SSIM) to quantitatively show the minimum measurable source intensity and the impact of stray light on both systems. Furthermore, we exploit the nonlocality inherent in entangled photon sources to remotely adjust the polarization state incident on the sample, which allows us to extract birefringence information without any interference on the measurement arm.

Unlike the aforementioned experimental work on quantum polarimetry, we perform a two-dimensional measurement for an inhomogeneous material compared to the single-spot polarization measurement performed in Refs (20-22). Moreover, compared to the ghost birefringence imaging shown in Ref (23), we further provide a detailed comparison between quantum and classical 2D polarimetric imaging to intuitively demonstrate the high sensitivity and robustness of the quantum system in the low illumination regime. The introduction of the Senarmont compensation method and the improvement of the four-step phase shift method allow us to obtain a more precise measurement for the phase retardation measurement and to extend the detectable optical axis range from $[0,\pi/2]$ to $[0,\pi]$, respectively. Our work paves the way for exploring potential applications in the measurement of photosensitive materials, active biological samples and other remote monitoring scenarios.

## Results

The experimental setup of the quantum polarimetric imaging system is shown in Fig. 1. It consists of three main parts: An entangled photon source part for the generation of polarization-entangled photon pairs based on the Sagnac interferometer configuration (24) and the spontaneous parametric down conversion (SPDC) of a type-II phase-matched periodically poled $KTiOPO_4$ (PPKTP) crystal. A sample measurement part that allows photon-sample interaction and collecting the polarimetric information using the raster scanning method (15). Two polarization analysis parts used for each of the separated photon pairs in both the upper control path and the lower measurement path for polarization base selection and joint measurement.

Before analyzing unknown samples using the above experimental setup, we initially assess the entangled photon source and calibrate our quantum polarimetric imaging system using a commercially available quarter wave plate (QWP). The conventional four-step phase shift method is introduced and further improved to calculate the phase retardance and optical axis angle in all the following experiments. After careful calibration, the system can achieve a nanometer-level accuracy in phase retardance measurements and ensuring an accuracy of ±0.1° for the angle of optical axis, feasible for imaging the 2D inhomogeneous anisotropic samples in the following sections. Further details of the experimental configuration and the calculation methods can be found in the **Materials and Method section**.

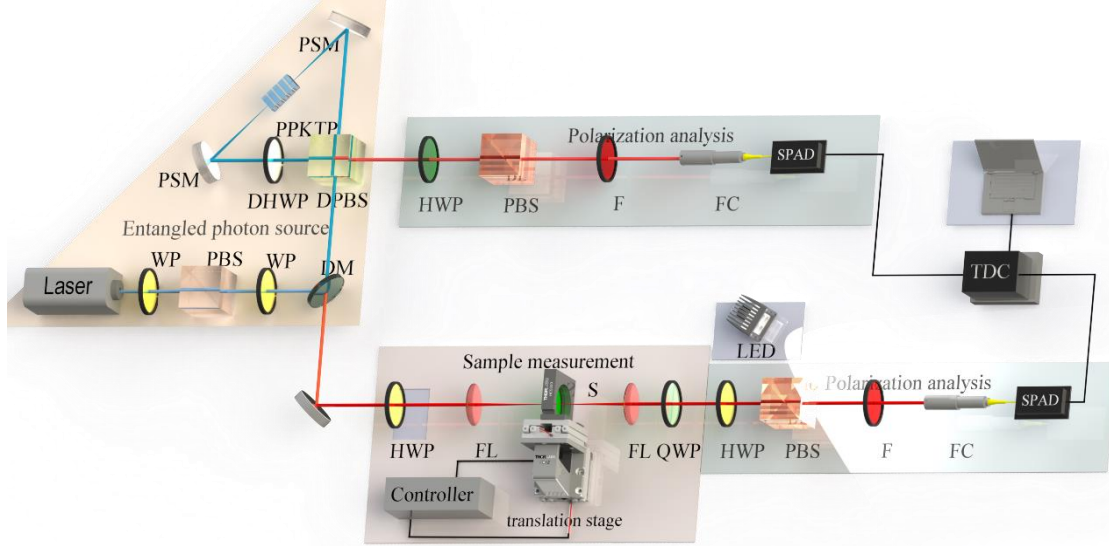

**Fig.1 Experimental setup. Entangled photon source part:** WP, waveplate set, including a quarter wave plate and a half wave plate (HWP); PBS, polarizing beam splitter; DM: dichroic mirror; DPBS, dichroic PBS; DHWP, dichroic HWP; PSM, off-axis parabolic silver mirror; PPKTP, periodically poled KTP crystal; **Sample measurement part:** Motorized translation stage; FL, focal lens; S, sample. **Polarization analysis part:** F, filter; FC, fiber collimator. SPAD, Single-photon Avalanche Diode. TDC, Time-to-digital converter.

**Polarimetric imaging under low illumination**

In this section, the sample is a polymer depolarizer characterized by a periodic birefringence pattern. Under conditions of extremely low illumination, we record the phase retardances and optical axis orientations obtained by both quantum and classical imaging setups. The comparative outcomes are shown in Fig.2.

Fig.2 illustrates the phase retardance and optical axis angle measurements for the polymer depolarizer across an imaging area of $2.5\,\text{cm} \times 2.5\,\text{cm}$. Utilizing a 0.25 cm incremental movement in two orthogonal directions, we capture a total of $11 \times 11$-pixel images for each measurement. The left two columns present the results from a quantum imaging system using local operations, specifically by rotating the HWP right after the sample in the measurement arm, while maintaining a fixed position for the HWP on the control side to determine the polarization state of the input photons. The HWP in the deep purple box of Fig.1 is temporarily removed for these measurements. The middle two columns display the results from a classical imaging system. Notably, the transformation of the measurement arm into a classical polarimeter is achieved by inserting an additional HWP (as indicated in the deep purple box) ahead of the focal lenses, thereby altering the polarization of the input photons. Consequently, variations in the single photon count at different HWP orientations provide valid information for the sample's birefringence properties.

Comparing the measurement results depicted in Fig.2, it is evident that both systems exhibit comparable trends and periodic pattern for the tested sample. However, as the input intensity decreases, the classical system shows a noticeable deviation from the original result. From $2 \times 10^4$ cps to $2\times10^3$ cps, the maximum and minimum values remain the same, while the width of the peaks gradually increases. Upon further reduction of the input intensity to $1\times10^3$ cps, the phase retardance value appears a significant deviation from its initial number. And the periodic pattern completely disappeared for both phase retardance and optical axis angle when the input intensity reduces to $5\times10^2$ cps. In contrast, the quantum system consistently demonstrates high fidelity in both shape and value from $2\times10^4$ cps to $5\times10^2$ cps. To assess the performance of both systems, we have employed the SSIM (25) as a metric. The calculated results, as illustrated in the bottom inset of Fig.2, reveal that the quantum system sustains an SSIM value above 0.8 across all tested illumination levels. Conversely, the classical system's SSIM progressively declines to 0.13 as the input intensity decreases, thereby vividly highlighting the superiority of the quantum polarimetric imaging system under low-illumination conditions.

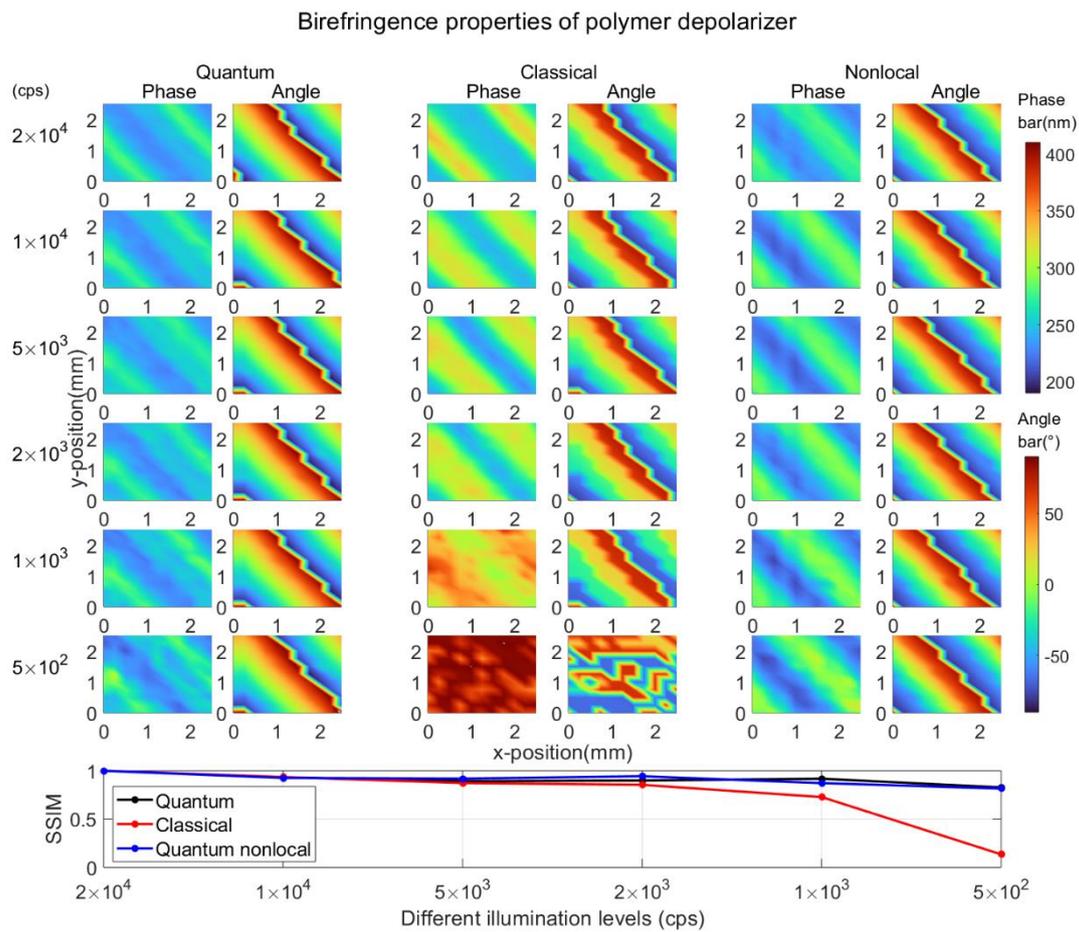

**Fig.2 Phase retardance and optical axis angle under different input illumination**

The left numbers show the input photon numbers for the images in the same row. The left, middle and right two columns show the phase retardance and the optical axis angle measured by the quantum imaging system with local operation, the classical imaging system and the quantum system with nonlocal operation, respectively.

The bottom sub-figure shows the SSIM of each phase retardation image.

**Nonlocality**

As a unique feature of entangled photon pairs, we also show the nonlocality of photons when measuring sample birefringence. The corresponding results under varying illumination conditions are presented in the right two columns of Fig.2. Compared with the experimental configuration depicted in Fig.1, we have relocated the QWP compensator to the control arm and adjusted the angle of HWP within the control arm for each measurement. Consequently, without any changes in the measurement arm, we are able to obtain the phase retardance and optical axis angle of the sample located in the measurement arm by only adjusting the components in the control side.

The results of the quantum nonlocal test show similar trends to those of the quantum local experiment. As evidenced by the images in Fig.2, the phase retardance and the optical axis angle exhibit very little sensitivity to diminishing illumination levels. Moreover, the remote-control capability offers distinct advantages over the quantum local experiment by significantly minimizing the disturbance to the measurement arm. This feature suggests promising applications in specialized environments, including but not limited to the remote sensing and biomedical diagnostics.

**Polarization imaging under perturbance**

In this section, we introduce an interference light source to investigate the robustness of quantum polarimetric imaging under conditions of perturbation. A light-emitting diode (LED) in the measurement arm, as depicted in Fig.1, operates in three distinct illumination modes: weak illumination (50lm), strong illumination (150lm) and intermittent flashing mode. In strong illumination mode, the LED can introduce approximately $2\times10^3$ cps of additional stray photons for single arm detection. In order to introduce enough perturbance, we decrease the measurement illumination to the same level, i.e. $1\times10^3$ cps, to compare the performance of two systems. The comparative analysis of the systems' performance is presented in Fig.3.

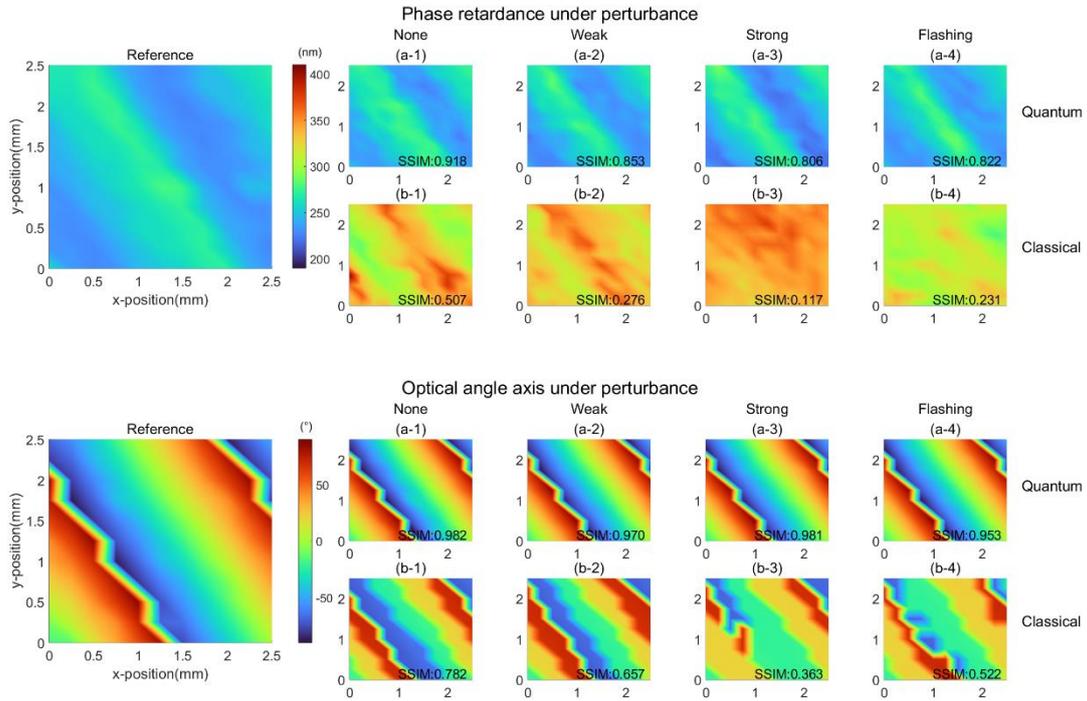

**Fig.3. Phase retardance and optical axis angle under perturbance.**

The left subplot (Reference) presents the measurement outcomes under $2\times10^4$ cps conditions without any interference. In contrast, the remaining subplots depict measurements conducted at a $1\times10^3$ cps illumination level.

**(a-1) to (a-4)**: Illustrate the phase retardance and the optical axis angle obtained by the quantum system, under conditions of no perturbation, weak perturbation, strong perturbation, and flashing perturbation, respectively.

**(b-1) to (b-4):** Obtained by the classical system under the same conditions as quantum system.

Fig.3 presents the birefringence property variations of the polymer depolarizer under three distinct types of perturbation. The imaging area, which remains at $2.5\text{cm} \times 2.5\text{cm}$ with a total of $11 \times 11$ pixels, differs from the area analyzed in the previous section. The first row of images corresponds to the results obtained from the quantum polarimetric imaging system. It is observed that all four images exhibit a high degree of consistency. The SSIM values, indicated at the lower corners of each image, are all above 0.8 for the three perturbation conditions, signifying excellent agreement with the reference.

In contrast, the second row displays the results from the classical system, which exhibit variability in both values and patterns. Notably, the measurement under zero perturbation already deviates from the reference, consistent with the result shown in the row $1 \times 10^3$ cps of Fig.2. The presence of perturbation further degrades the detection outcomes. Specifically, the SSIM values for the phase retardances decline to 0.276 for weak perturbation, 0.117 for strong perturbation, and 0.231 for flashing perturbation, indicating a significant reduction in similarity to the reference image.

Given that the perturbations examined so far have not yet revealed the limitations of the quantum system, we have conducted an additional investigation focusing solely on the quantum system. Due to the fixed mode intensity of the interference source, which prevents us from further increasing the perturbance level, we have opted to reduce the intensity of the photon source instead. This approach allows us to further compare the measurement results between an unperturbed environment and a strongly perturbed scenario for the quantum system.

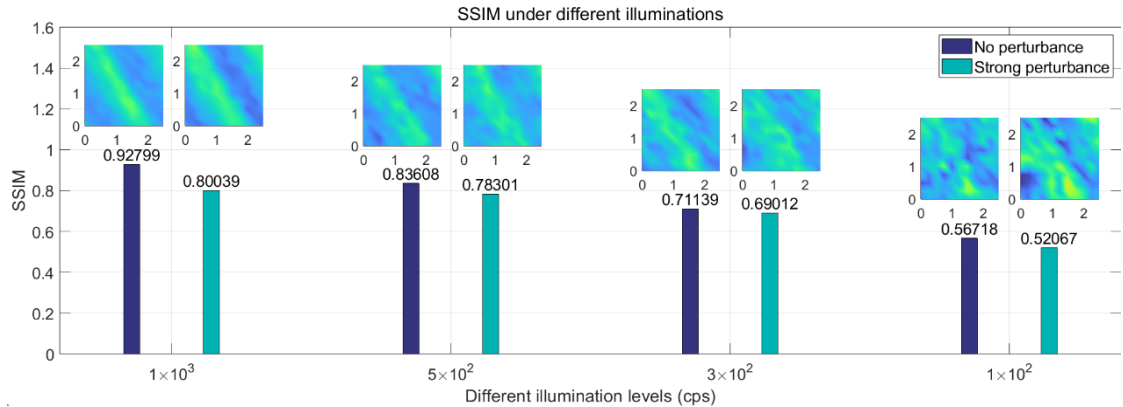

Fig.4 Comparison of SSIMs in unperturbed and strong-perturbed environments

The left bar in each group represents the SSIM values obtained under unperturbed conditions for $1 \times 10^3$, $5 \times 10^2$, $3 \times 10^2$, $1 \times 10^2$ cps input illuminations, respectively. Right bar of each group shows the SSIM with strong perturbance under the same illuminations.

Fig.4 illustrates that the periodic distribution characteristic of the sample diminishes as the input illumination decreases to 100 photons per second. Notably, the strong perturbation at 150lm has a negligible impact on the measurements obtained by the quantum polarimetric system across all input levels. This resilience can be attributed to the dual-path structure of the quantum system, where the coincidence photon detection mechanism demonstrates a high degree of perturbance resistance due to the presence of an unperturbed control arm. In fact, even if the control arm is subjected to interference, provided that the interference originates from two distinct sources, the system will maintain a remarkable level of robustness against external perturbations.

## Discussion

In summary, we build a quantum polarimetric imaging system to measure the birefringence properties of anisotropic samples with high precision under low illumination conditions. By combining an improved four-step phase shift method with the Senarmont compensation technique, our system achieves nanometer-level accuracy in phase retardance measurement and $\pm 0.1°$ for the optical axis angle measurement. Taking a periodically distributed polymer depolarizer as a sample, we compare the image quality between the quantum system and its classical counterpart over a range of decreasing illumination levels. Besides, we also investigate the impact of different perturbation modes on the obtained images and explore the unique nonlocal property of the quantum system.

The results show that quantum system has higher sensitivity, enabling it to accurately detect samples even at significantly lower input illumination levels, down to as few as 100 photons per second. As the input illumination gradually reduces from $2\times 10^4$ cps to $5\times 10^2$ cps, the SSIM for each measurement in the quantum system remains consistently high, exceeding a similarity of 0.8. In contrast, the classical system experiences an obvious decline in performance at $1\times 10^3$ cps and retrieve to 0.13 with a $5\times 10^2$ cps input. Furthermore, the quantum system demonstrates enhanced robustness against stray light compared to the classical system. It can feasibly resist perturbations from various interference modes, ranging from strong (150lm) to weak (50lm) perturbance, with only minimal degradation ($\leq 0.1$) in SSIM. While the classical system suffers a deterioration of 0.39 in SSIM for strong perturbance and 0.23 for weak perturbance.

Our experiment highlights the quantum polarimetric imaging system with its dual-way structure and coincidence measurement, which efficiently shields against background noise and perturbations affecting the single-arm. This design significantly enhances the signal-to-noise ratio, greatly improving the sensitivity and robustness of the birefringence measurements. The excellent performance in low illumination field opens up a wide range of unique applications. It provides a delicate method for detecting photosensitive materials and monitoring fragile bioprocesses without the risk of photobleaching or thermal damage, benefiting the fields of biometric diagnosis and material growth. Also, the nonlocality of the quantum system greatly reduces the susceptibility to disturbances in the measurement arm. The inherent high resistance to single-arm perturbations satisfies the need for the various subfields of remote sensing.

## Materials and Methods
### Experimental configuration

A 405nm continuous wave diode laser (Kunteng Quantum Technology Co. Ltd) with a 45° linear polarization enters the central dichroic polarizing beam splitter (DPBS) of the Sagnac interferometer. The split horizontal and vertical polarized light pass clockwise and counterclockwise through the loop, which is formed by two symmetrical parabolic silver mirrors (PSM) and a 45° tilted dichroic half-wave plate (DHWP). Each of the bidirectional light pumps a type-II phase-matched PPKTP crystal (Raicol Crystals Ltd.) to generate the orthogonally polarized photon pairs by SPDC. Then the central DPBS recombines the correlated photon pairs to form the polarization-entangled photon source.

In the sample measurement part, a two-dimensional motorized translation stage is used to move the sample precisely with a minimal step size of $0.2\mu m$. Two plano-convex lenses with focal lengths of 50mm focus the light onto the sample surface. The raster scanning technique, although more time consuming, is used for the 2D imaging because of the higher signal-to-noise ratio (SNR) compared with the full-field imaging technique (26). After the sample measurement part, we use a QWP as a compensator to improve the performance of the system, the benefit of which has been analyzed in our previous work (27).

Both the control and measurement arms have the same configuration of collection components, i.e. a true zero-order half wave plate (HWP) for base selection, an 808 nm PBS for projective measurement, a long pass filter and a bandpass filter to remove the pump light. The emitted photons are then collected in a single-mode fiber and detected by a single-photon avalanche diode (Excelitas, SPCM-AQRH) with a detection efficiency of approximately 60% at a wavelength of 810 nm.

**Calculation method**

We use the improved four-step phase-shifting method to derive the sample's birefringence information from coincidence photon counts. Following the experimental setup in Fig.1, the output coincidence counts for the joint measurement of control and measurement arms are:

$$\begin{aligned} N &= N_0 \left| tr\left( \hat{E}_c \otimes \hat{E}_m \left( \hat{J}_c \otimes \hat{J}_m \right) |\phi\rangle_{cm} \langle\phi| \right) \right|^2 \\ &= \frac{N_0}{4} \{ 1 - \sin\delta \sin(4h_m - 2q)\sin(4h_c + 2\theta) + \cos(4h_m - 2q) \\ &\quad \times [-\cos(2q - 2\theta)\cos(4h_c + 2\theta) + \cos\delta \sin(2q - 2\theta)\sin(4h_c + 2\theta)] \}. \end{aligned} \quad (1)$$

Here, the subscripts $c$ and $m$ stand for the control and measurement arm, while $h$ and $q$ represent the angle of HWP and QWP, respectively. $|\phi\rangle_{cm} = 1/\sqrt{2}[|HV\rangle + |VH\rangle]$ is the initial entangled photon state generated by the Sagnac interferometer. $\hat{J}_{c(m)}$ is the evolution operator of the photons in the control (measurement) arm. Based on the experimental setup, we have $\hat{J}_c = I$ and $\hat{J}_m = \hat{U}_{QWP}(q)\hat{U}_{sample}(\delta, \theta)$, where $\delta$ and $\theta$ stand for the phase retardance and optical axis angle of the tested sample. Base selection and projective measurement are contained in the projective operator $\hat{E}_{c(m)}$ with HWP angle set to $h_{c(m)}$ in the control (measurement) arm.

Based on Eq. (1), the birefringence parameters $\delta$ and $\theta$ can be calculated by varying the controllable $h_{c(m)}$ to get its corresponding coincidence counts. The conventional four-step phase-shifting method selects four angles for the analyzer, i.e., 0°, 45°, 90° and 135°, to derive the birefringence parameters. However, the range of the detectable optical angle $\theta$ is usually limited to $[0, \pi/2]$. To meet the requirements of our anisotropic sample, which has an optical axis angle varying from 0 to $\pi$, we improve the conventional four-step phase-shifting method by performing the method twice in both H-base ($h_c = 45°$) and D-base ($h_c = 22.5°$) in the quantum polarimetric imaging system. In this way, we are able to extend the measurable range from $[0, \pi/2]$ to $[0, \pi]$. Further calculation details are showed in the **Supplementary Materials**.

**Acknowledgement**
Sincerely thanks for the technical support of Suzhou PTC Optical Instrument Co., Ltd.
**Fundings**：This work is supported by the National Key Research and Development Program of China (2022YFB3607700, 2022YFB3903102), National Natural Science Foundation of China

# Supplementary Materials

**Calculation Process of Improved Four-step Phase-shifting Method**

Here we show the improved four-step phase-shifting method combined with Eq. (1). When the HWP in the control arm is fixed at 45° while the compensated QWP is fixed at 0°(H-basis), the output photon numbers under four degrees are:

$$IH_{1(h_m=0°)} = \frac{1}{8}(3+\cos 4\theta + 2\cos\delta\sin^2 2\theta), \quad IH_{2(h_m=22.5°)} = \frac{1}{4}(1+\sin\delta\sin 2\theta),$$

$$IH_{3(h_m=45°)} = 2\cos^2\theta\sin^2\frac{\delta}{2}\sin^2\theta, \quad IH_{4(h_m=62.5°)} = \frac{1}{4}(1-\sin\delta\sin 2\theta). \tag{S.1}$$

Therefore, we have:

$$AH = \frac{IH_1 - IH_3}{IH_1 + IH_3} = \cos^2\frac{\delta}{2} + \cos 4\theta \sin^2\frac{\delta}{2},$$

$$BH = \frac{IH_2 - IH_4}{IH_2 + IH_4} = \sin\delta\sin 2\theta. \tag{S.2}$$

The corresponding phase retardance and optical axis can be derived from the above parameters:

$$\delta_H = \arccos\left[\frac{1-BH^2-AH}{AH-1}\right], \quad \theta_H = \frac{1}{2}\arcsin\left(\pm\sqrt{\frac{(AH-1)^2}{2-2AH-BH^2}}\right). \tag{S.3}$$

Note that the measurable phase retardance range is confined to $[0,\pi]$, we could decide the sign before the square root in Eq. (S.3) based on the sign of BH in the Eq. (S.2). In this situation, we have $\theta_H \in [-\pi/4, \pi/4]$.

Similarly, when the HWP in the control arm is fixed at 22.5° while the compensator at 45° (D-basis), the output photon numbers under four degrees are:

$$ID_{1(h_m=0°)} = \frac{1}{4}(1+\sin\delta\cos 2\theta), \quad ID_{2(h_m=22.5°)} = \frac{1}{4}(1+\cos\delta\cos^2 2\theta + \sin^2 2\theta),$$

$$ID_{3(h_m=45°)} = \frac{1}{4}(1-\sin\delta\cos 2\theta), \quad ID_{4(h_m=62.5°)} = \frac{1}{2}\cos^2 2\theta\sin^2\frac{\delta}{2}. \tag{S.4}$$

Therefore, we have:

$$AD = \frac{ID_1 - ID_3}{ID_1 + ID_3} = \sin\delta\cos 2\theta,$$

$$BD = \frac{ID_2 - ID_4}{ID_2 + ID_4} = \cos^2\frac{\delta}{2} - \cos 4\theta \sin^2\frac{\delta}{2}. \tag{S.5}$$

The corresponding phase retardance and optical axis can be derived from the above parameters:

$$\delta_D = \arccos\left[\frac{1-AD^2-BD}{BD-1}\right], \quad \theta_D = \frac{1}{2}\arccos\left(\pm\sqrt{\frac{(BD-1)^2}{2-2BD-AD^2}}\right). \tag{S.6}$$

By combining Eq. (S.2) and (S.5), the sign of $\sin 2\theta$ and $\cos 2\theta$ can be determined. Therefore, the measurable range of the optical axis angle $\theta$ is extended to $[0,\pi]$. Besides, the phase retardance is calculated from:

$$\delta = \arccos[AH + BD - 1]. \tag{S.7}$$

## Photon source evaluation

The performance of the entangled photon source is evaluated by temporally removing the sample from the measurement arm. Two-photon polarization interference curves are collected under H-basis (*hc* fixed at 0°) and D-basis (*hc* fixed at 22.5°). The raw interference visibilities (V=(C$_{max}$-C$_{min}$)/(C$_{max}$+C$_{min}$))) are calculated to be 99.75% ± 0.01% and 99.12% ± 0.01% for H- and D-basis, respectively, well above the 71% indicated by the Bell inequality. In addition, the CHSH inequality is measured to quantify the entanglement quality. The S-parameter is calculated to be 2.754 ± 0.006191 in 10 s, violating the CHSH inequality by 122 standard deviations. With a 2mW pump input, the maximum coincidence counts can achieve $2\times10^4$ cps, with total collecting efficiencies of the control and measurement arms to be 13.68% and 14.23%, respectively.

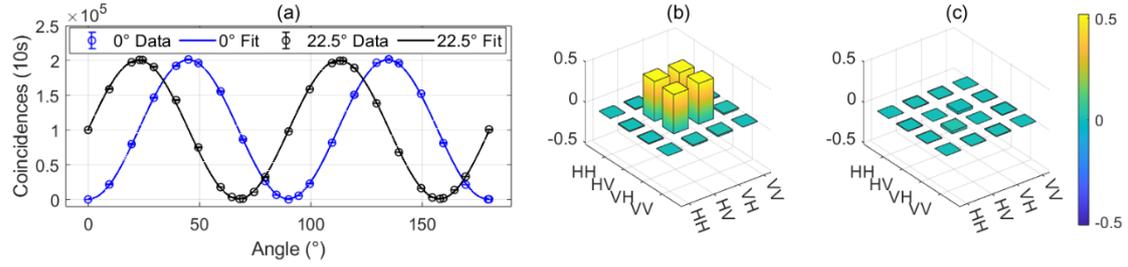

**Fig.S1 Entangled photon source properties.**

**(a)** two-photon polarization interference curve; **(b)&(c)** real and imaginary part of reconstructed density matrix.

## System calibration

Based on the above entangled-photon source, we measure a commercial QWP (808 nm, MFOPT Ltd.) to calibrate the angle of each polarization-correlated optical component and test the measurement accuracy of the system.

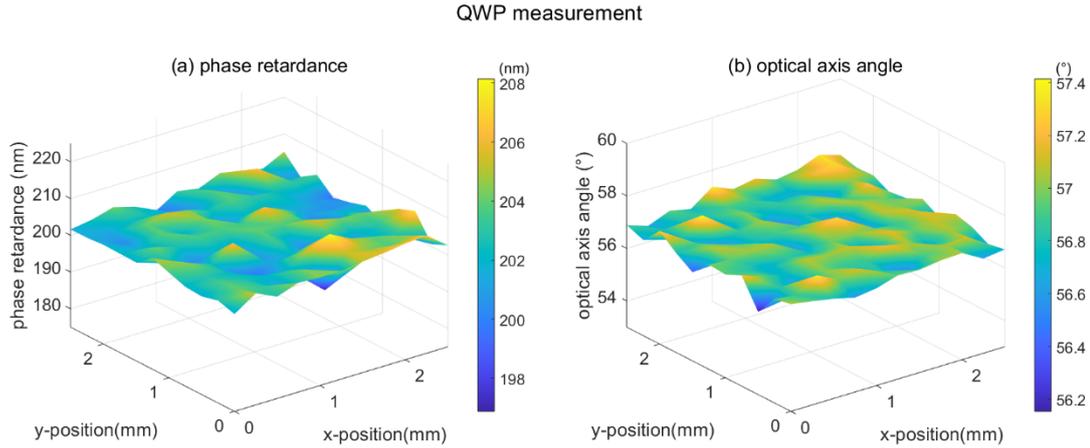

**Fig.S2 Measurement result of a standard QWP. (a)** phase retardance and **(b)** optical axis angle

Fig.S2 presents the measurement results for a QWP with a central wavelength of 808 nm. The dimensions of the imaging area are $2.5\,\text{cm}\times2.5\,\text{cm}$, with each incremental movement of translation stage set at 0.25cm. The average phase retardance, along with its standard deviation across $11\times11$ pixels, is measured to be 202.88±0.51nm, closely aligning with the theoretical prediction of $\lambda/4$. Additionally, the average optical axis angle and its standard deviation are recorded as 56.9±0.06°. These results indicate that the precision of our quantum polarimetric imaging system is capable of achieving nanometer-level accuracy in phase retardance measurements and ensuring an accuracy of ±0.1° for angle measurement.